\documentclass[useAMS,usenatbib]{mn2e}
\usepackage{graphicx}
\usepackage{graphics}
\usepackage{amssymb}
\usepackage{amsmath}
\usepackage{algpseudocode}
\usepackage{multicol}
\usepackage{longtable}
\usepackage{algorithm}
\usepackage{array}
\usepackage{rotating}

\title[Reduction of astrochemical networks]{Complexity reduction of astrochemical networks}

\author[T. Grassi et al.]{T. Grassi$^{1,2}$, S. Bovino$^{1}$, F. A. Gianturco$^{1}$\thanks{Corresponding author: fa.gianturco@caspur.it}, P. Baiocchi$^{3}$, and E. Merlin$^{2}$\\
$^{1}$Department of Chemistry, Sapienza University of Rome, P.le A. Moro, 5, 00185 Roma\\
$^{2}$Department of Astronomy, Universit\`a degli Studi di Padova, Vicolo dell'Osservatorio, 3, 35122 Padova\\
$^{3}$Department of Mathematics ``G. Castelnuovo'', Sapienza University of Rome, P.le A. Moro, 5, 00185 Roma}

\begin{document}
\newcommand{\ith}{$i$th }
\newcommand{\jth}{$j$th }

\newcommand{\dd}{\mathrm d}
\newcommand{\mA}{\mathrm A}
\newcommand{\mB}{\mathrm B}
\newcommand{\mC}{\mathrm C}
\newcommand{\mD}{\mathrm D}
\newcommand{\mE}{\mathrm E}
\newcommand{\mH}{\mathrm H}
\newcommand{\mSi}{\mathrm Si}
\newcommand{\mO}{\mathrm O}
\newcommand{\cmc}{\mathrm{cm}^{-3}}
\newcommand{\real}{\mathbb R}
\newcommand{\superscript}[1]{\ensuremath{^{\scriptscriptstyle\textrm{#1}\,}}}
\newcommand{\trader}{\superscript{\textregistered}}

\newcommand\mnras{MNRAS}
\newcommand\apj{ApJ}
\newcommand\aap{A\&A}
\newcommand\osu{osu\_01\_2007}

\date{Accepted 2012 June 18.  Received 2012 June 18; in original form 2012 April 30}

\pagerange{\pageref{firstpage}--\pageref{lastpage}} \pubyear{2012}

\maketitle

\label{firstpage}

\begin{abstract}
We present a new computational scheme aimed at reducing the complexity of the chemical networks in astrophysical models, one which is shown to markedly improve their computational efficiency. 
It contains a flux-reduction scheme that permits to deal with both large and small systems. This procedure is shown to yield a large speed-up of the corresponding numerical codes and provides good accord with the full network results. We analyse and discuss two examples involving chemistry networks of the interstellar medium and show that the results from the present reduction technique reproduce very well the results from fuller calculations. 
\end{abstract}


\section{Introduction}
Chemistry is one of the most important ingredients in many astrophysical simulations (e.g. \citealt{GalliPalla98}, \citealt{Nelson1999}, \citealt{Maio07}, \citealt{MerlinChiosi07}, \citealt{Tesileanu2008}, \citealt{Glover2010}, \citealt{Gnedin2009}, \citealt{Yamasawa2011}).
Unfortunately chemical processes, usually represented through a network, consist of a large ensemble of reactions that tightly connect a given set of chemical species. From a mathematical point of view this network is represented by a system of ordinary differential equations (ODE) that often constitutes a system of stiff, coupled equation.
Computationally speaking, it means that a lot of the cpu-time is spent to solve the chemical network evolution instead of solving, for instance, the hydrodynamical equations. 

To deal with this problem there are various techniques which can be applied. The most used is probably the direct reduction approach that consists in pre-selecting the most important species and then their most important reactions for a certain astrophysical environment (e.g. \citealt{Nelson1999}). This method is trivial when the network is small, or when there are small regions of the network that are easy to visualize and to isolate, but when the complexity grows this approach could result in large errors, since the network reduction is based on \emph{ad hoc} chemical or physical considerations applied only once before solving the ODEs.

Another widely used approach is to reduce the number of non-equilibrium\footnote{Not to be confused with non-LTE.} species (e.g. \citealt{Glover2010}). Under this assumption the differential equations are explicitly solved only for some selected species while the remaining species are taken to reach instantaneously their equilibrium. Even this reduction technique is strongly \mbox{problem-dependent} and this approximation could lead to large uncertainties in the final abundances.

The aim of the present work is to suggest instead a method that increases the efficiency of the system of differential equations which describes the chemical network, but without introducing any \emph{a priori} assumption, while providing an approximate but still accurate solution that matches that from the full network calculation.

The general method was developed by \cite{Tupper2002} (hereafter TUP02) while we introduce modifications to fit our class of astrophysical scenarios. We shall show that the TUP02 scheme with our improvements yields accurate results while achieving large computational speed-ups. We shall apply the method to a small chemical network (similar to \citealt{Glover2010}) and to a larger one (similar to \citealt{Wakelam2008}).
We first introduce the mathematical problem in Section \ref{chemical_network}, then we describe in Section \ref{our_approach} the main features of the present scheme and show our modifications and upgrades to the original method. Finally, Section \ref{astrophysical_scenarios} discusses the chosen astrophysical examples. 

\section{Earlier network reducing methods: an outline}\label{chemical_network}
A chemical network with $N$ species is represented by a system of $N$ ordinary differential equations as
\begin{equation}\label{ode}
	\frac{\dd n_i}{\dd t} \equiv \dot n_i = -  n_i \sum_r k_{ir} n_r + \sum_p\sum_q k_{pq} n_p n_q	\,,
\end{equation}
where $i\in\real^N$, $n_i$ is the abundance\footnote{Abundances are always considered as number densities unless otherwise indicated.} of the \ith species\footnote{A species can be both an element or a molecule. e.g. H$_2$ and Si are both species.} and $k_{ip}$ is the reaction rate that represents the probability for the reaction between the \ith and the \jth species  to occur at a chosen temperature. The right-hand side (RHS) of Eq.(\ref{ode}) is divided into two parts: the first is a sum that describes the reactions which destroy the \ith species while the second part is a sum over the reactions that produce the \ith species. Here $i$, $p$, $q$, and $r$ indicate each a specific species.

Eq.(\ref{ode}) can be now written as
\begin{equation}\label{ode_flux}
	\dot n_i = \sum_{j=1}^M s_{ij}F_j(n,k)\,,
\end{equation}
where $s_{ij}$ is an element of the stoichiometric matrix, \mbox{$F_j=k_{lm}n_ln_m$} is a flux, and $M$ is the total number of reactions. The elements of the stoichiometric matrix are integers and can be $s_{ij}=\pm1$ (depending on the direction of the reaction) and $s_{ij}=0$ when the reaction does not occur. The flux is the amount of mass or number density that moves from a species to another in a unit of time. Equation (\ref{ode_flux}) allows to viewing the network as a \emph{directed graph}, where each node (or vertex) represents the number density of each species, and each link (or edge) represents a flux. The direction (and the existence) of an edge is determined by the corresponding element of the stoichiometric matrix $s$. Note that here $i$ indicates a species, while $j$ now represents a flux.

This system can be easily solved with DVODE \citep{Hindmarsh83} using a backward differentiation formulas (BDF) integration scheme. The computational cost of solving the ODE system is given by the evaluation of the RHS of the Eq.(\ref{ode_flux}) and in some cases also by the large number of equations. Hence we need to find a way of reducing the complexity of the system. There are two possible approaches to this end: one is based on reducing the number of differential equations and another keeps the number of equations constant but reduces the number of elements in the RHS of each equation.
These methods are not mutually exclusive.

The first technique to reduce the system consists in reducing the number of species through some lumping method (e.g. \citealt{Okino98}) like the singular value decomposition. In this case the full system - see Eq.(\ref{ode_flux}) - becomes a system of $K<N$ equations
\begin{equation}\label{hat_ode_flux}
	\frac{\dd\hat n_i}{\dd t} = \sum_{j=1}^{\hat M} \hat s_{ij}\hat F_j(\hat n,k)\,,
\end{equation}
where $i\in\mathbb{R}^K$. The hat indicates the same quantities of Eq.(\ref{ode_flux}) but transformed by a unitary $N\times K$ matrix $U$ that permits $\real^N\overset{U}{\longrightarrow}\real^K$ and analogously its transposed $U'$ lets $\real^N\overset{U'}{\longleftarrow}\real^K$. This method allows to solve the ODE system in a different space where $\hat n\in\real^K$ are linear combinations of the various $n\in\real^N$ and one chooses the transformation $U$ in order to obtain a simpler system. 
Unfortunately the main drawback of species-based reduction is that building the transformation matrix $U$ may not be trivial and also the computational cost to build $U$ could easily result in a computational overhead.

In the present work we therefore follow a different approach that consists in reducing the fluxes between the species while keeping the same number of differential equations. In other words, we look for a system with $W<M$ represented by the following ODE system
\begin{equation}\label{ode_flux_reduced}
	\dot n_i = \sum_{j=1}^W s_{ij}F_j(n,k)\,,
\end{equation}
where $i\in\real^N$, but in this case the sum runs over fewer elements than those of the system of Eq.(\ref{ode_flux}). The advantage of this method is that the greater part of the computational cost is given by the RHS of Eq.(\ref{ode_flux}) and we reduce the number of reactions instead of reducing the number of species. In particular, we delete the reactions that are \emph{slow} or, in other words, the reactions that have small fluxes.
This method has some practical advantages compared to the one which is species-based: (i) the mass conservation is always guaranteed because we are removing whole reactions, and (ii) the selection of small fluxes is more efficient than building a unitary transformation, thus decreasing the computational overhead.
On the other hand, the drawback of this method is that it requires a robust selection criterion to determine what are the important reactions. 

A combined species-based and reaction-based methods has been suggested by \citet{Wiebe2003} and we focus our attention on their reaction-based method because it is more interesting for the present discussion. They search for the production and destruction reactions that are most important for the evolution of an important species, and then they assign their own relative significance. The most important species are determined by an iterative algorithm that computes an importance weight for each species to determine what reactions are important or not. Their interesting method led to good results but unfortunately it seems to be problem-dependent and, moreover, it is affected by unexpected changes of the physical conditions of the evolving model. The latter is the most troublesome drawback.

The method proposed by \cite{Tupper2002} belongs to the class of the reaction-based methods:
the main assumption is to consider only the fluxes that are larger than a given default value, namely $\varepsilon$, in order to select the fluxes only if $\varepsilon<\left|F_i\right|$, this procedure being repeated at each \emph{macro-step}, instead of doing it once for all at the beginning of the simulation. It is called macro-step to underline the fact that is larger than the usual integration steps, so that a macro-step includes one or more integrations steps.
This is based on the approach of \cite{Petzold1999} and already described by Eq.(\ref{ode_flux_reduced}), although TUP02 introduces some differences, the most important of which consist of using different reduced models over short periods of time instead of a single reduced system over the entire integration interval. This approach turns out to be more effective although problems may arise since: (i) one needs to determine the size of a macro-step and (ii) the method requires an error control to check the validity of the macro-step integration.

To solve the first problem TUP02 considers the influence of the omitted reactions to predict for how long the model will remain valid. In particular one needs to know the influence of the omitted reactions over the next macro-step or, even better, when such reactions will become large enough to make the chosen macro-step not valid.
TUP02 uses finite differences to compute the rate of change of a single reaction obtaining $G_j\approx\dd F_j(n)/\dd t$. Then he considers that the reaction most likely to cause a macro-step rejection is the one with the largest $G_j$, this assumption permitting to find the size of a macro-step (\citealt{Tupper2002}, Section 4.3).

The second problem depends on the omitted reactions: if they become too large (larger than a default tolerance) during a single macro-step, the integration over that time period is not valid. To ensure the effectiveness of the reduction TUP02 suggests to compare the omitted reactions at the beginning and at the end of the macro-step. If there is any difference between these two check-points the macro-step is rejected and then the entire integration is repeated including the omitted reactions. This procedure is iterated until the difference disappears. Our present scheme, however, does not use any rejection technique because we use a more accurate macro-step, as will be discussed in Section \ref{our_approach}, so that for our models we found that a rejection scheme is not necessary.  

\cite{Tupper2002} then applies the method to a large astrophysical network and to a small combustion network. For the first case the speed-up is very large, and the differences of results between the full model and the reduced one are negligible. In the combustion case the full system is well reproduced, but it becomes computationally inefficient since a large overhead occurs, so that the reduction method is slower than the straightforward integration. It shows that a massively interconnected system could be difficult to reduce, in particular when all the fluxes have the same absolute values.  
On the other hand, in the first example, the algorithm easily finds a subset of reactions that is dominant compared to the others.
\section{The present reduction scheme}\label{our_approach}
Our approach is similar to the one proposed by \cite{Tupper2002}, although we introduce the following important modifications:

\textsc{1. Adaptive tolerance:} we define the tolerance $\varepsilon$ relative to  the largest flux as $\varepsilon = \zeta \left|F_\mathrm{max}\right|$. This relative tolerance changes during the evolution of the system, depending on the magnitude of the largest flux, the one that dominates the model. We then select the fluxes that obey to the following rule
\begin{equation}\label{condition}
	\left|F_j\right|>\varepsilon = \zeta \left|F_\mathrm{max}\right|\,,
\end{equation}

where $\zeta$ depends on the desired accuracy (the best value for $\zeta$ will be discussed later).

\begin{figure}
\begin{center}
	\includegraphics[width=.3\textwidth]{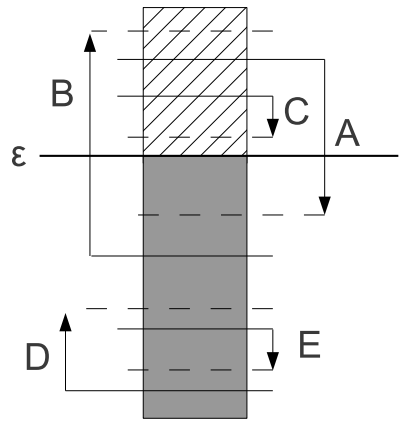}
	\caption{The evolution of fluxes in a macro-step. See text for further details.}
	\label{flux_speed}
\end{center}
\end{figure}

\textsc{2. Macro-step length:} another modification to the scheme of \cite{Tupper2002} is the choice of the macro-step length. We first consider the aforementioned equation \mbox{$G_j\approx\dd F_j(n)/\dd t$}. This equation can be viewed as the velocity of the \jth flux: in other words, it shows how much the flux of a given reaction changes during a time interval. We use this quantity to predict when a reaction will cross the value $\varepsilon$. Figure \ref{flux_speed} represents the amount of different fluxes, with the line labelled $\varepsilon$ indicating the tolerance value. The fluxes in the hatched area above that line obey the condition (\ref{condition}), while the fluxes below the $\varepsilon$-line are not considered in the reduced system. Each horizontal solid line is the flux value at a given time $t$, and the corresponding dashed line is the value of the same flux at time $t+\Delta t$, as indicated by the arrow which represents the variation of flux over $\Delta t$; the larger the $G_j$ the longer the arrow. We show five different cases: (A) a reaction important at time $t$ becomes uninteresting after $\Delta t$, since it crosses the $\varepsilon$-line to reach the darker area, (B) a reaction becomes important after $\Delta t$, (C) a reaction for which the contribution remains important, and (D, E) two reactions which remain unimportant even if one of them increases its importance in a macro-step. To estimate the length of a macro-step we are interested in the reactions that belong to the  
classes (A) and (B), since their fluxes change their significance before the end of a macro-step. We calculate the macro-step length as
\begin{equation}\label{macrolength_abs}
	\Delta t_\mathrm{new} = \phi\,\cdot\,\mathrm{min}\left\{\frac{\left|F_j^{t} -\varepsilon\right|}{\left|F_j^{t} -F_j^{t+\Delta t}\right|+\varphi}\Delta t\right\}\,,
\end{equation}
where $(F_j^{t} -\varepsilon)$ represents the \jth flux distance from the threshold $\varepsilon$, and $(F_j^{t} -F_j^{t+\Delta t})$ is the flux variation over the previous macro-step of length $\Delta t$. A small value $\varphi$ is introduced to avoid divisions by zero. The constant factor $\phi\ge1$ is the time-tolerance, a parameter to avoid excessively short macro-steps caused by the reactions that oscillate around the threshold $\varepsilon$. Note, however, that if $\phi$ is chosen to be too large it can affect the accuracy of the method (in this paper we use $\phi=10$).

\begin{figure}
\centering
	\includegraphics[width=.3\textwidth, angle=-90]{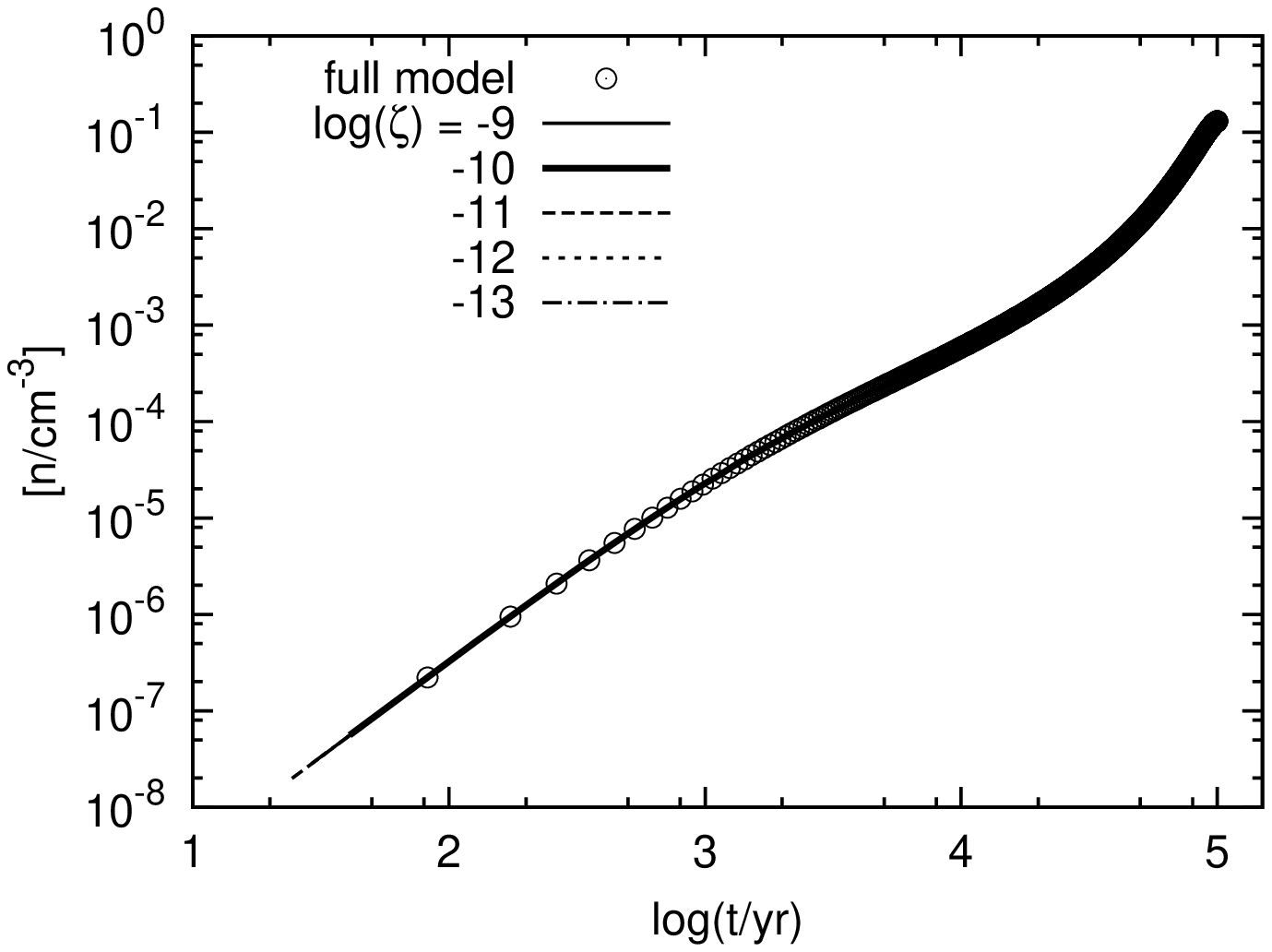}
	\includegraphics[width=.3\textwidth, angle=-90]{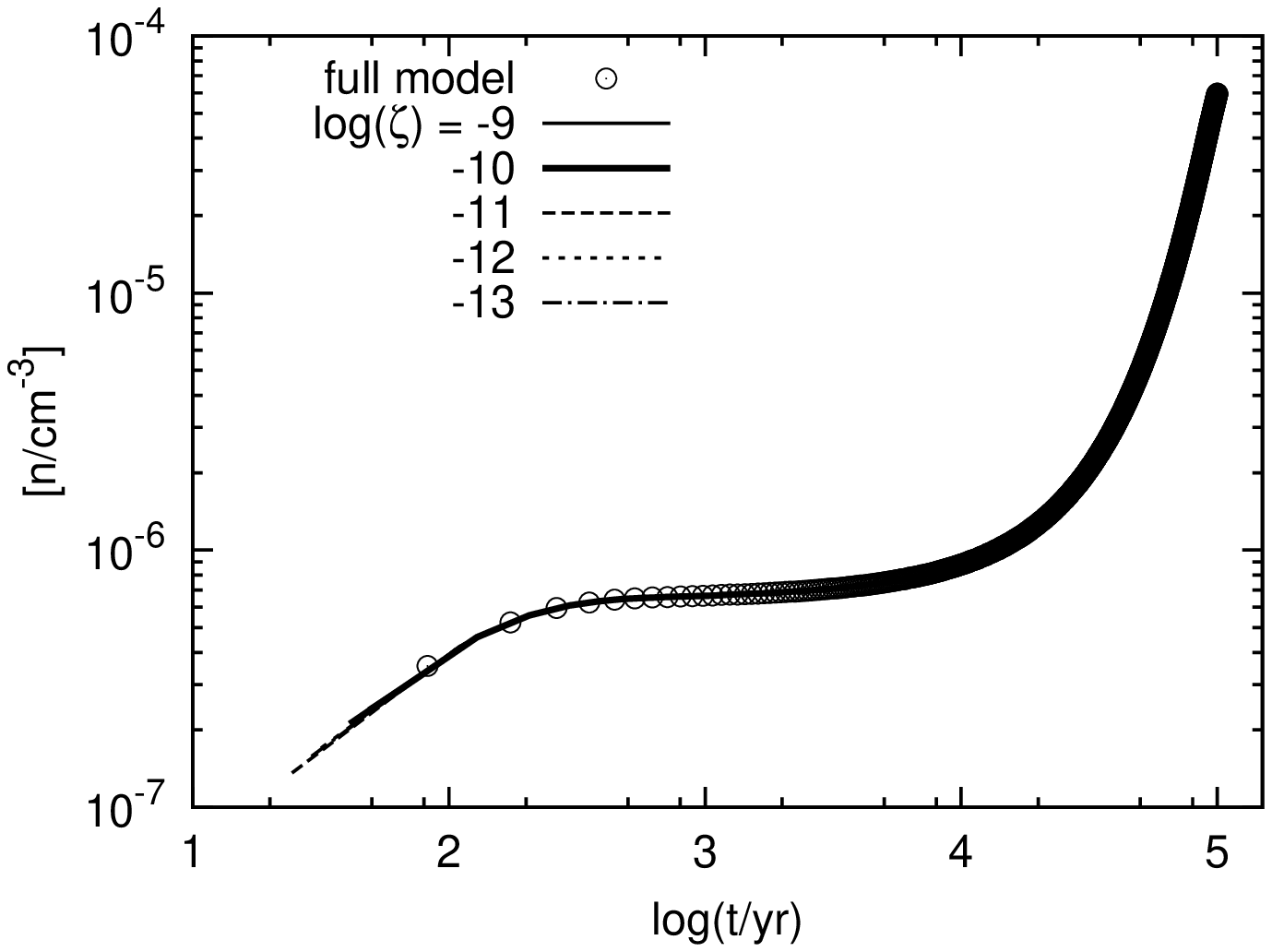}
	\caption{Small model calculations. Time evolution of C$^+$ (top) and OH (bottom) number densities. Open circles indicate the full model evolutions, while lines are for different $\zeta$ in the model. All lines are clearly overlapping.}\label{gloC2_H2O}
\end{figure}

It is worth noticing that the direction of the flux velocity is important, and then fast growing reactions that are above the threshold should be ignored when one determines the length of a macro-step.  This consideration is also true for the fast decreasing reactions that are smaller than $\varepsilon$. To take into account this feature and to avoid shorter, unnecessary macro-step lengths in Eq.(\ref{macrolength_abs}), one should write it as
\begin{equation}
	\Delta t_\mathrm{new} = \phi\,\cdot\,\mathrm{min}\left\{\frac{F_j^{t} -\varepsilon}{F_j^{t} -F_j^{t+\Delta t}}\Delta t\right\}\,,
\end{equation}
although for programming purposes we shall still use Eq.(\ref{macrolength_abs}).
The reader must be aware that this macro-step calculation is based on a linear growth hypothesis (see $G_j$) and it could be used together with a rejection scheme, as done by \cite{Tupper2002}, if the system presents a strong non-linear behaviour. Again it should be mentioned that for our modelling we found it to be unnecessary.

\textsc{3. Sorting:} We further improved on the TUP02 method  by introducing a flux-sorting device at each macro-step. The system of Eq.(\ref{ode_flux}) is built with a loop over all the $M$ reactions, and each reaction is evaluated with the condition (\ref{condition}), hence we must evaluate this condition $M$ times: this operation has a non-negligible computational cost. If the fluxes are sorted out, then we can break the loop after $W$ iterations - see Eq.(\ref{ode_flux_reduced}). This strategy is not effective if the macro-steps become very small, because the time to sort the fluxes becomes larger than the time gained in the early loop-breaking. In the astrophysical tests employed in the present paper we never incurred in a computational overhead, even if we used a simple bubble sorting. One should note here that the first sorting is the most expensive while the others are very efficient, because they find an almost-sorted array. This reduces the inefficiency of the bubble sorting procedure \citep{Knuth1997}. To increase the sorting efficiency we introduce a boolean array where each element is false or true depending on  condition (\ref{condition}). We sort then the boolean array instead of the array that contains the fluxes because we are only interested in evaluating the condition (\ref{condition}) rather than generating sorted fluxes. 

The overall efficiency of our approach is determined by the time spent to prepare a macro-step $\tau_\mathrm{m}$ compared to the solver's integration time over a macro-step. For a macro-step $\Delta t_\mathrm{macro}$ we must have
\begin{equation}\label{efficient}
	\tau_\mathrm{m} < \tau_\mathrm{do} \frac{\Delta t_\mathrm{macro}}{\Delta t_\mathrm{ode}}=\tau_\mathrm{do}\cdot \mathcal{C}\,,
\end{equation}
where $\tau_\mathrm{do}$ is the time to calculate a single do-cycle-step, $\Delta t_\mathrm{ode}$ is an integration step of the ODE solver. The last ratio determines the number of ODE solver's calls in a macro-step to the function that computes the RHS of Eq.(\ref{ode_flux}).
We roughly assume that without sorting the cost of creating a macro-step is given by $\tau_\mathrm{m}=3M$ (namely $M$ to find the maximum, $M$ to create the boolean array, and $M$ to compute the length of the macro-step, where $M$ is defined in Eq.(\ref{ode_flux}) in the previous Section) when introducing the bubble sorting, the worst possible case is $\tau_\mathrm{m}=3M+M^2$. Analogously, the reduced do-loop needs $\tau_{do}=W\tau_\mathrm{dn}+M\tau_\mathrm{if}$ steps to calculate the RHS of our differential equations, where $W$ is defined in Eq.(\ref{ode_flux_reduced}), $\tau_\mathrm{dn}$ is the cost of evaluating $\dd n$, and $\tau_\mathrm{if}$ is the cost of evaluating the condition (\ref{condition}). When we introduce sorting we have $\tau_\mathrm{if}=0$. The efficiency is then guaranteed if $\mathcal{C}>3\,M/(W\tau_\mathrm{dn}+M\tau_\mathrm{if})$ and if $\mathcal{C}>(3M+M^2)/(W\tau_\mathrm{dn})$ with and without sorting, respectively. It is worth to mention at this point that using sorting for our examples enabled us to obtain $\tau_\mathrm{sorted} \approx0.75\,\tau_\mathrm{unsorted}$.

We illustrate our scheme by briefly outlining a pseudo-code (see Algorithm \ref{pseudocode}), where $n$ is an array containing the abundances of the species, $t$ is the simulation time, $t_\mathrm{macro}$ is the length of a macro-step, $t_\mathrm{targ}$ is the target time to solve the system, $t_\mathrm{end}$ is the length of the simulation, $\varepsilon$ is the flux limit, $\zeta$ is defined in Eq.(\ref{condition}), $F$ is the array of the fluxes, and $u$ is the aforementioned boolean array.

\begin{algorithm*}\caption{ - Pseudo-code of the scheme proposed in this work. See text for further details.}
\label{pseudocode}
\begin{multicols*}{2}
	\begin{algorithmic}
		\State $n\gets n_0$
		\State $t\gets 0$
		\State $t_\mathrm{macro}\gets 0.01$
		\State $t_\mathrm{targ}\gets t_\mathrm{macro}$
		\Repeat
			\State $\varepsilon\gets \zeta\cdot\max(F^t)$
			\State $u\gets \mathrm{evaluate}(\varepsilon,\,F^t)$
			\State $[u,F]\gets\mathrm{sort}(u)$
			\State $t_\mathrm{macro}\gets \mathrm{get\_macrostep\_length(\Delta F,\,\Delta t)}$
			\State $t_\mathrm{targ}\gets \max(t+t_\mathrm{macro},\,t_\mathrm{end})$
			\State $\mathrm{odesolve}(t,\,t_\mathrm{targ},\,n,\,F,\,u)$
		\Until{$(t<t_\mathrm{end})$}
	\end{algorithmic}
\columnbreak

\begin{algorithmic}
	\State //initialize $n$
	\State //initialize $t$
	\State //set a default value for the first macro-step
	\State //set the solver target time
	\State //loop over the macro-steps
	\State //compute flux limit
	\State //make boolean array
	\State //sort $u$ and $F$ using $u$
	\State //compute the length of the next macro-step
	\State //set next target time
	\State //solve the reduced system for $[t,t_\mathrm{targ}]$
	\State //exit when the end is reached

\end{algorithmic}
\end{multicols*}
\end{algorithm*}

As an implementation of our scheme, this pseudo-code will be applied to two different astrophysical scenarios in the following Section.
\section{Results and discussion}\label{astrophysical_scenarios}
\subsection{The small network example}\label{small_eg}

We now apply our scheme to a one-zone chemical network using a set of reactions similar to the one in \cite{Glover2010}, involving 29 species and 170 reactions (see Tab.\ref{tabtfrac_glo1} for the complete list). The chosen subset of reactions has only an illustrative purpose since we are not interested in reproducing the physical behaviour of a real astrophysical environment, but we want only to test the reduction method proposed. Hence, this set of reaction is intended only to mimic an astrophysical behaviour. These considerations are also true for the large network example.

We also note that in \cite{Glover2010} thirteen of the original species are considered to be in instantaneous chemical equilibrium, while the remaining nineteen species follow the full non-equilibrium evolution. In our model none of the 29 species are assumed to be in instantaneous equilibrium so that we track the non-equilibrium evolution for all of them.
Our initial conditions are $n=10^{-20}\,\cmc$ for all the species except $n_\mH=10^3\, \cmc$, $n_\mO=3.16\times10^{-4}n_\mH$, and $n_\mC=1.41\times10^{-4}n_\mH$. We also set $A_v=10$, and $T=10^3$ K, then we let the system evolve until $t_\mathrm{end}=10^5$ yr. For our test we did not use any cooling or heating, hence $\dd T/\dd t=0$.

The program is serial and it is written in FORTRAN 90, compiled with Intel\trader Fortran Compiler 12.1\footnote{Optimization flag used: -O3, -unroll, -ip, and -ipo. For further details contact the authors.} on an Intel\trader Xeon\trader E5430. The solver used is the DVODE \citep{Hindmarsh83} with absolute tolerance $10^{-40}$ and relative tolerance $10^{-6}$.

\begin{figure}
\begin{center}
	\includegraphics[width=.3\textwidth, angle=-90]{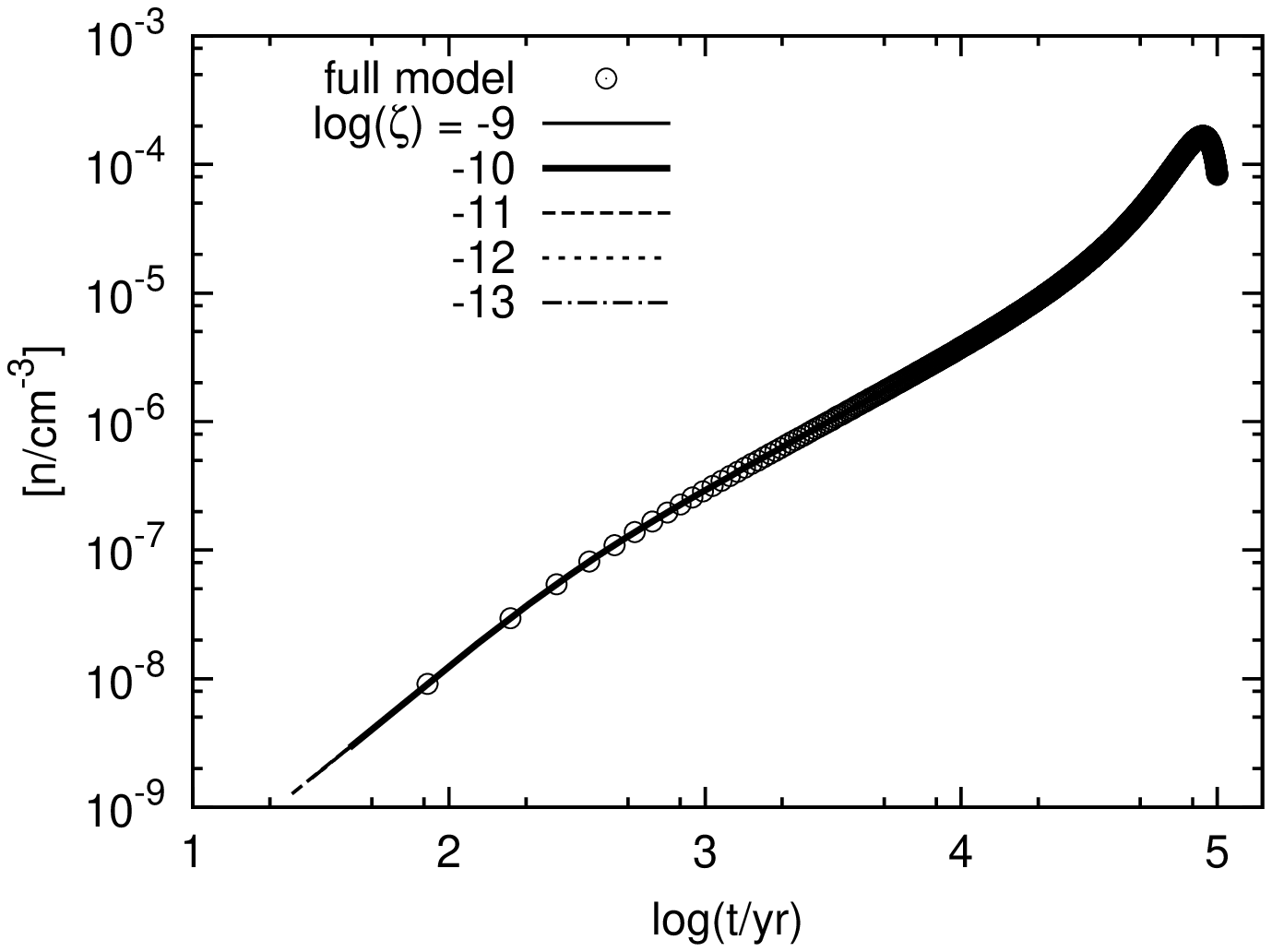}
	\includegraphics[width=.3\textwidth, angle=-90]{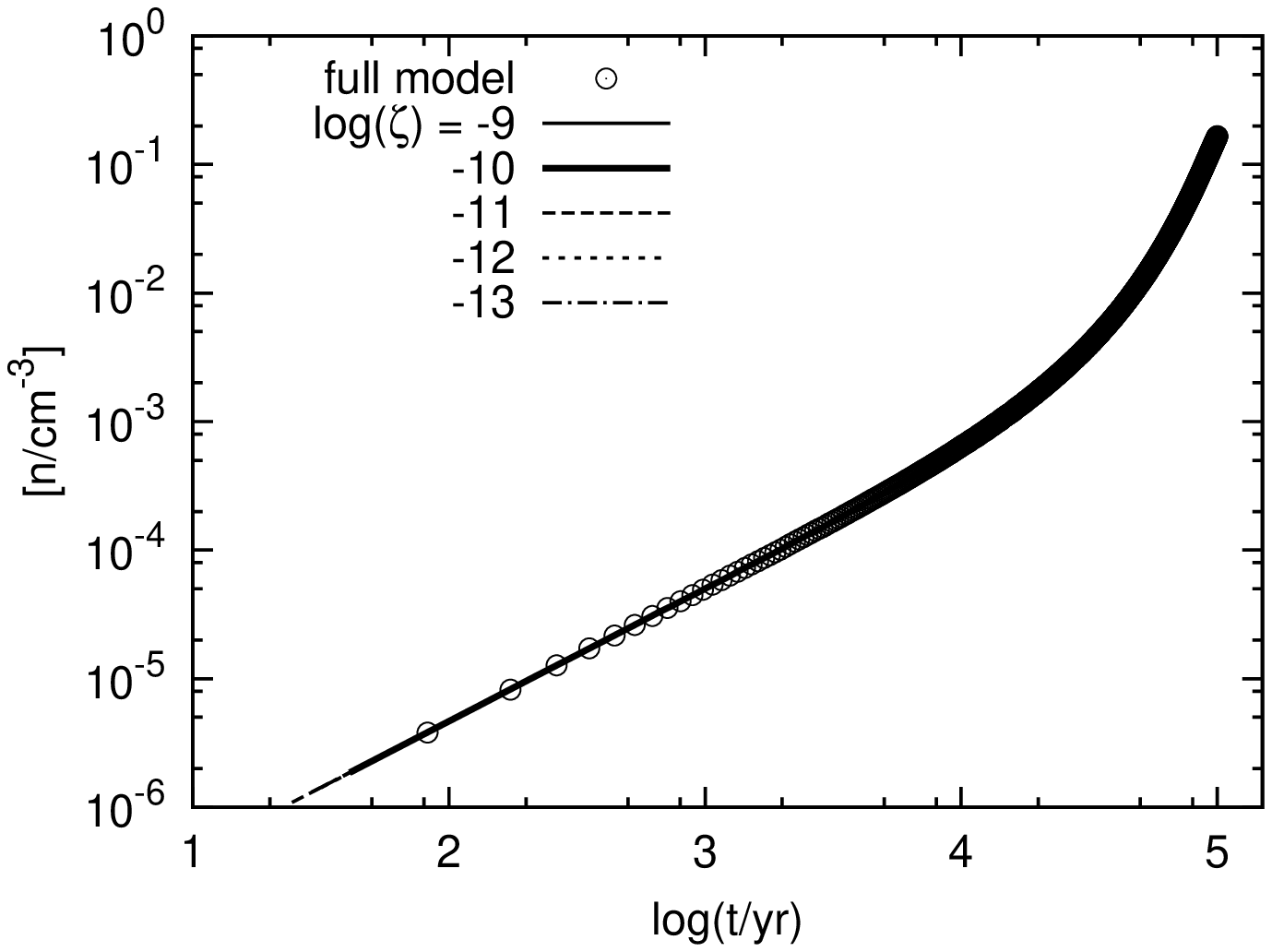}
	\caption{Small model calculations. Time evolution of CO (top) and H2 (bottom) number densities. Open circles indicate the full model evolutions, while the lines are for different $\zeta$. All lines are clearly overlapping.}
	\label{gloCO_H2}
\end{center}
\end{figure}

\begin{figure}
\begin{center}
	\includegraphics[width=.3\textwidth, angle=-90]{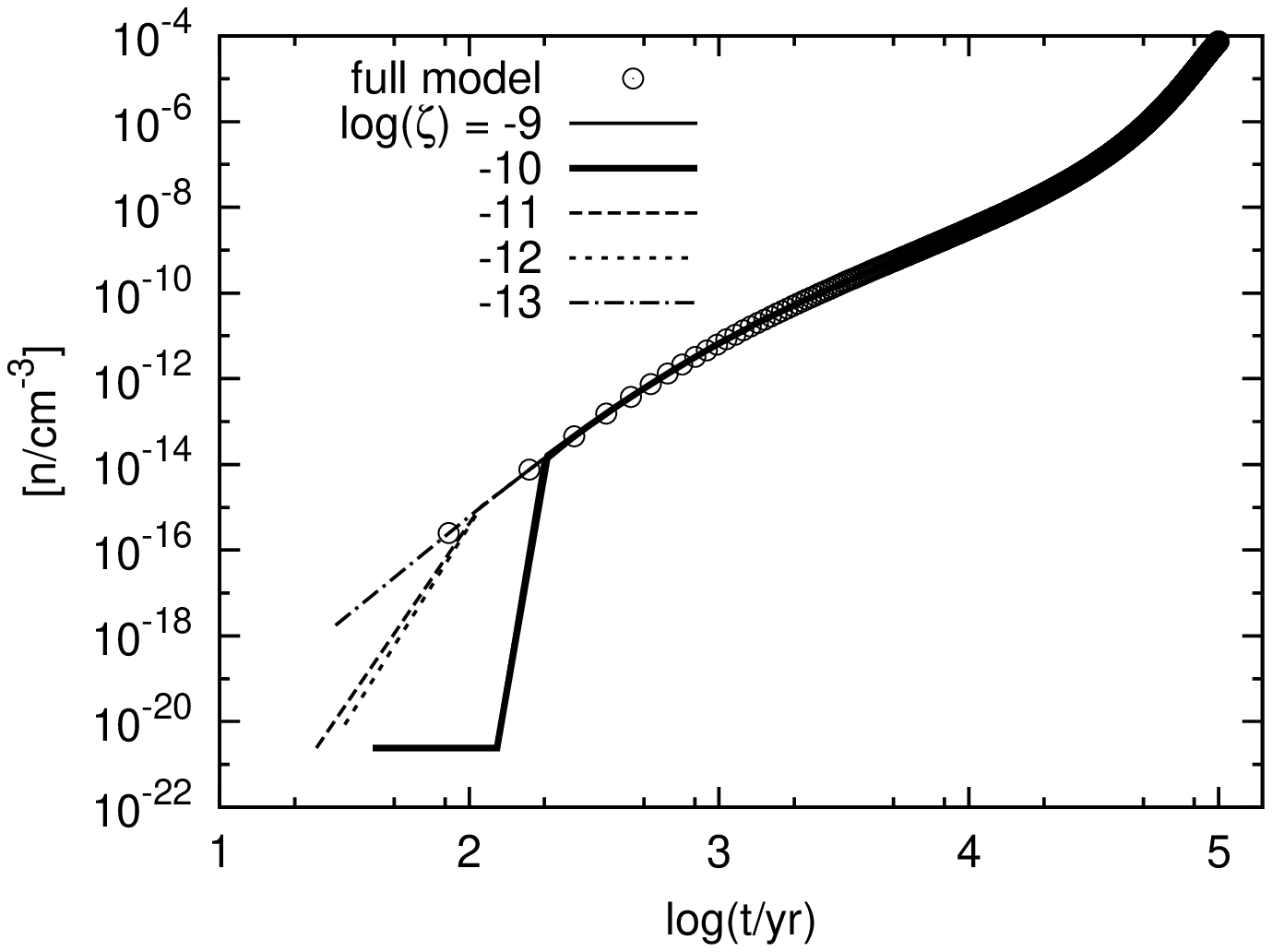}
	\includegraphics[width=.3\textwidth, angle=-90]{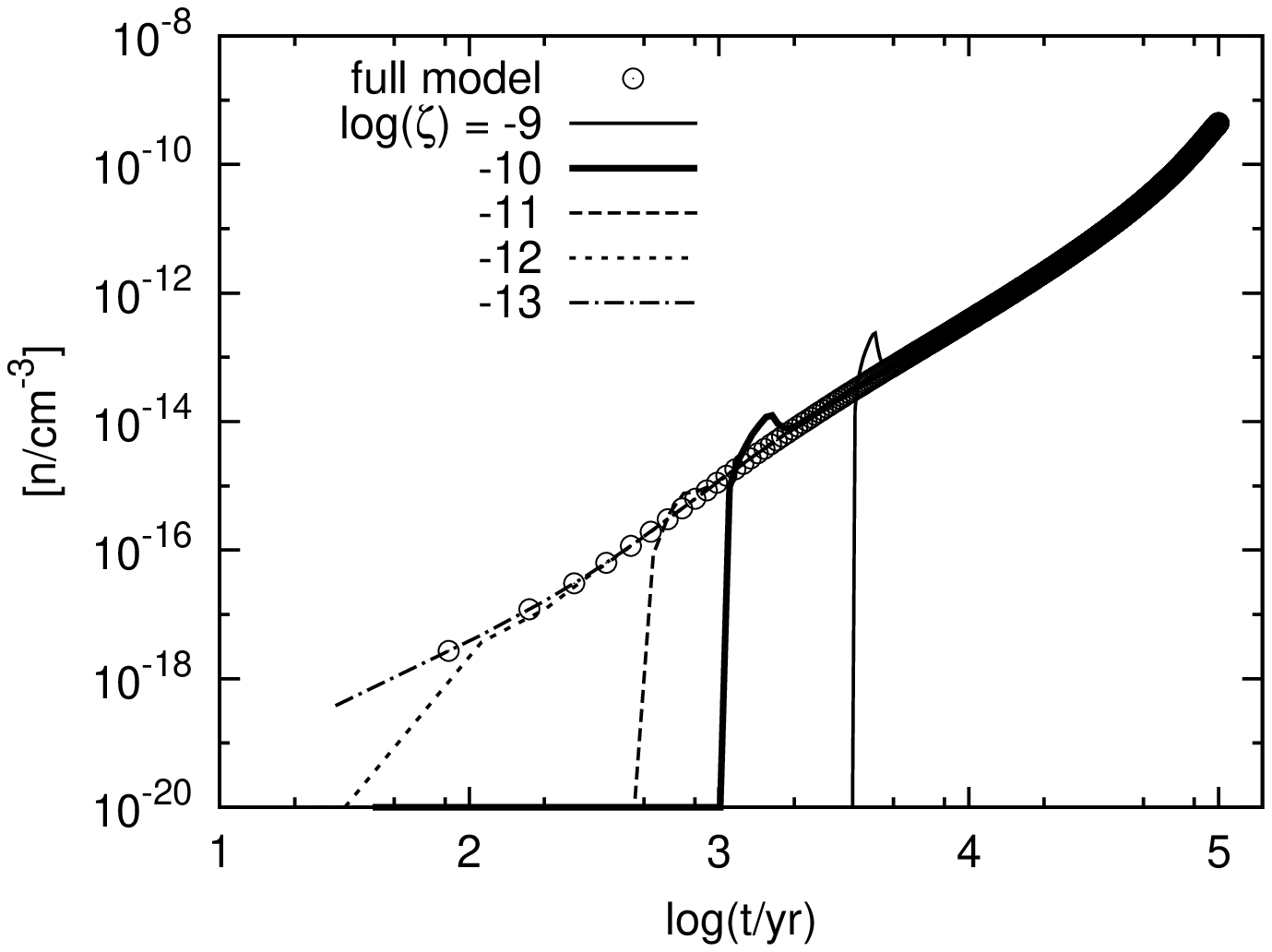}
	\caption{Small model calculations. Time evolution of H$^+$ (top) and CH$_2^+$ (bottom) number densities. Open circles indicate the full model evolutions, while lines are for different $\zeta$.}
	\label{gloHp_Op}
\end{center}
\end{figure}

We have made five simulations for five different values of $\zeta$. Our aim is to compare the time evolution of the various species provided by  the full model with the evolution from the reduced ones. The larger fraction of the species behaves as the examples shown in Fig.\ref{gloC2_H2O} and Fig.\ref{gloCO_H2}, where the full model is well reproduced by all the reduced models we employ. Some of the species evolve as in Fig.\ref{gloHp_Op}: in this case the full model starts to be well reproduced only when the importance of these species crosses a given value that depends on $\zeta$. We note that different reduced models eventually catch-up with the full model at different times, depending on the selected $\zeta$. It is important to remark that only the species with an overall small number density are affected by this behaviour.

\begin{figure}
\begin{center}
	\includegraphics[width=.3\textwidth, angle=-90]{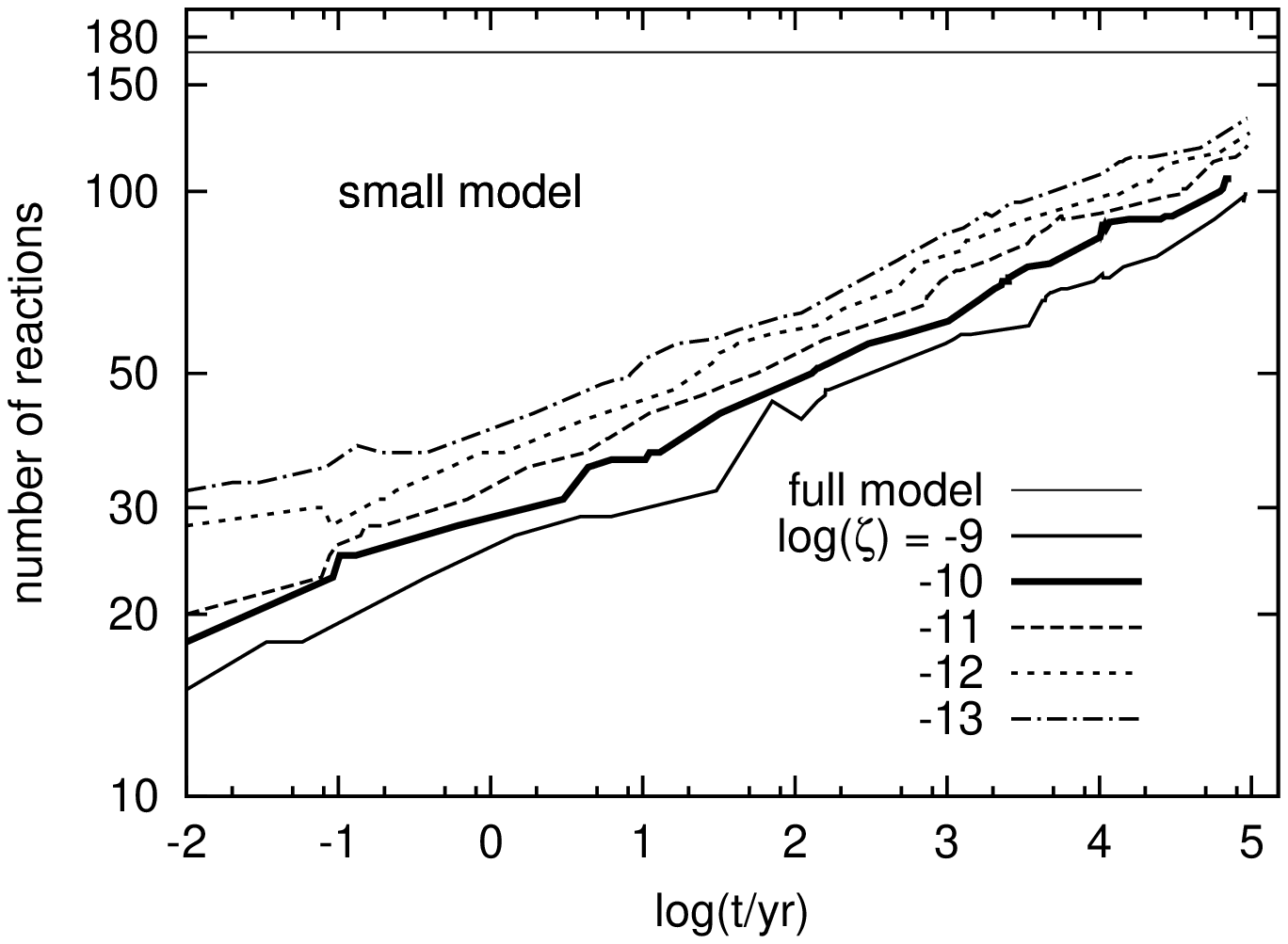}
	\includegraphics[width=.3\textwidth, angle=-90]{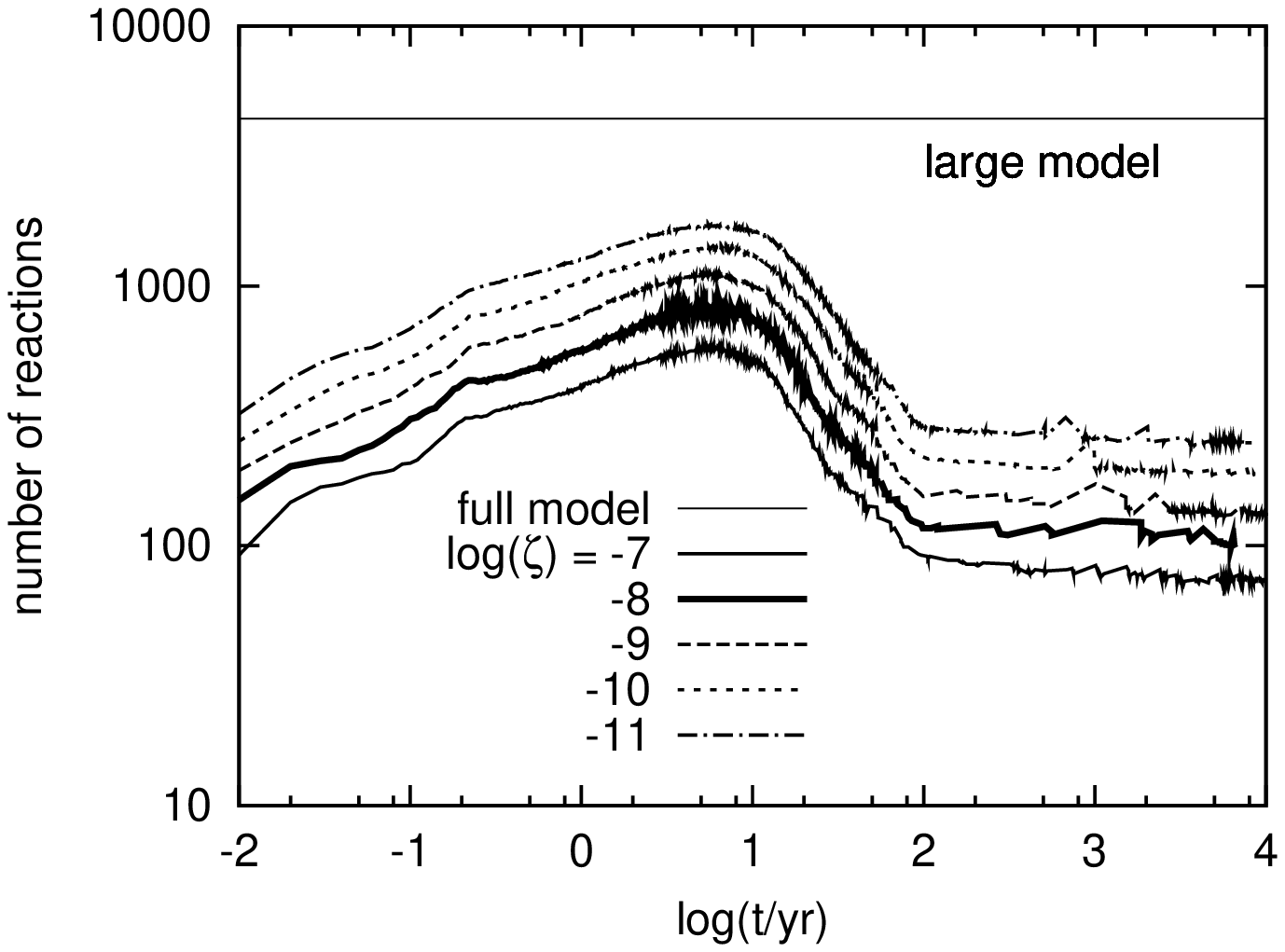}
	\caption{The time evolution of the reactions used in the reduced model with different $\zeta$ values. The horizontal line indicates the full model evolutions (constant reaction number), while varying lines are for different $\zeta$. Top panel: small model, bottom panel: large model. Note the logarithmic scale.}
	\label{glo_nrea}
\end{center}
\end{figure}

The computational speed-ups are indicated in Tab.\ref{glo_speed}. We also see that lower values of $\zeta$ imply slower integration times because, when the threshold is small, there are more reactions that affect the construction of the RHS of Eq.(\ref{ode_flux_reduced}). The number of reactions used at a given time is shown in the upper panel of Fig.\ref{glo_nrea}. For all the reduced models the number of reactions grows with time, because more reactions become important in the network when new species are formed. The number of reactions used at any given time is also proportional to $\zeta$, since more reactions cross the threshold $\varepsilon$, also proportional to $\zeta$. This is consistent with noting that, when $\zeta=0$, the reduced model corresponds to the full model and - analogously - when $\zeta\to0$ the reduced models converge to the full one. Hence, when $\zeta$ becomes very small the full model is almost exactly reproduced but a computational overhead would occur. Using Eq.(\ref{efficient}) it is possible to determine the value $\zeta_0$ that maximizes the accuracy without overheads, so when $\zeta<\zeta_0$ the reduction method is efficient and the choice of $\zeta$ depends on the desired accuracy/efficiency trade-off. These considerations suggest that the user must choose the value of $\zeta$ before doing the calculation, depending on the necessary accuracy and the efficiency. However, the tests proposed show that $\zeta=10^{-9}$ represents a good compromise.

Note that (Tab.\ref{glo_speed}) we obtain better timing when we introduce sorting, as expected, except for the large network when $\zeta=10^{-9}$. This unexpected result could be caused by a cache-locality effect which is architecture-dependent as proved by tests made on different machines (not shown here because the details are not relevant). It is also important to note that without using the sorting procedure, for the small network example with $\zeta=10^{-13}$, an overhead occurs. Conversely, introducing a sorting procedure allows not only to avoid such overhead, but also to obtain a large speed-up.

\begin{table}
	\caption{Speed-ups for different $\zeta$ and for small and large network examples, with and without sorting. Normalized to the full model.
	See the discussion in the text.}
	\centering
	\begin{tabular}{l|cc|cc}
		&\multicolumn{2}{c}{small}&\multicolumn{2}{c}{large}\\
		\hline
		$log(\zeta)$ & no sort & sort & no sort & sort\\
		\hline
		full & 1.00 & 1.00 & 1.00 & 1.00\\
		-7 & 0.71 & 0.52 & 0.07 & 0.05\\
		-8 & 0.78 & 0.58 & 0.08 & 0.06\\
		-9 & 0.81 & 0.61 & 0.08 & 0.08\\
		-10 & 0.85 & 0.65 & 0.16 & 0.11\\
		-11 & 0.88 & 0.68 & 0.20 & 0.14\\
		-12 & 0.99 & 0.71 & - & - \\
		-13 & 1.04 & 0.75 & - & -\\
	\end{tabular}
\label{glo_speed}
\end{table}

Using our present reduction method we can also check the importance of the different reactions in a given network by simply measuring the time that a chosen reactions is used by a reduced network. We compare these times with the total integration time $t_\mathrm{end}$ employed for the model with $\zeta=10^{-13}$ and we show the results for this comparison in the last column of Tab.\ref{tabtfrac_glo1}. Note that the first 113 reactions play a role in the model for more than $50\%$ of the total time, while the remaining reactions have a considerably smaller role over time. Table \ref{tabtfrac_glo1} is referred to the small network only.

\subsection{The large network example}
In this subsection we apply the present scheme to a chemical network similar to the one employed by \cite{Wakelam2008} - see the considerations made for the small example at the beginning of Sect.\ref{small_eg}. This network is less connected compared to the previous one, even if there are more species (451) and more reactions (4399). For such a system we expect to obtain a larger speed-up, because, due to its lower connectivity and then the possibility of deleting slower reactions, a set of almost-isolated subsystems could be identified. One should note that slow reactions are represented in our scheme by small fluxes, hence they can be ignored if they lie under the $\varepsilon$ threshold.
We follow the initial condition of the EA2 model in \cite{Wakelam2008} (except for $n_\mathrm{Si^+}=2.56\times 10^{-4}\,\cmc$), $n_\mathrm{H_2}=10^4\,\cmc$, while the remaining species are set to 
$10^{-20}\,\cmc$, and $n_\mathrm{e}=\sum{n_\mathrm{ion}}$. We let the system evolve until $t=t_\mathrm{end}=10^4$ yr, with $\zeta_\mathrm{CR}=1.3\times10^{-17}$ s$^{-1}$, $T=10$ K, and $\dd T/\dd t=0$. The model is dust-free.

\begin{figure}
\begin{center}
	\includegraphics[width=.3\textwidth, angle=-90]{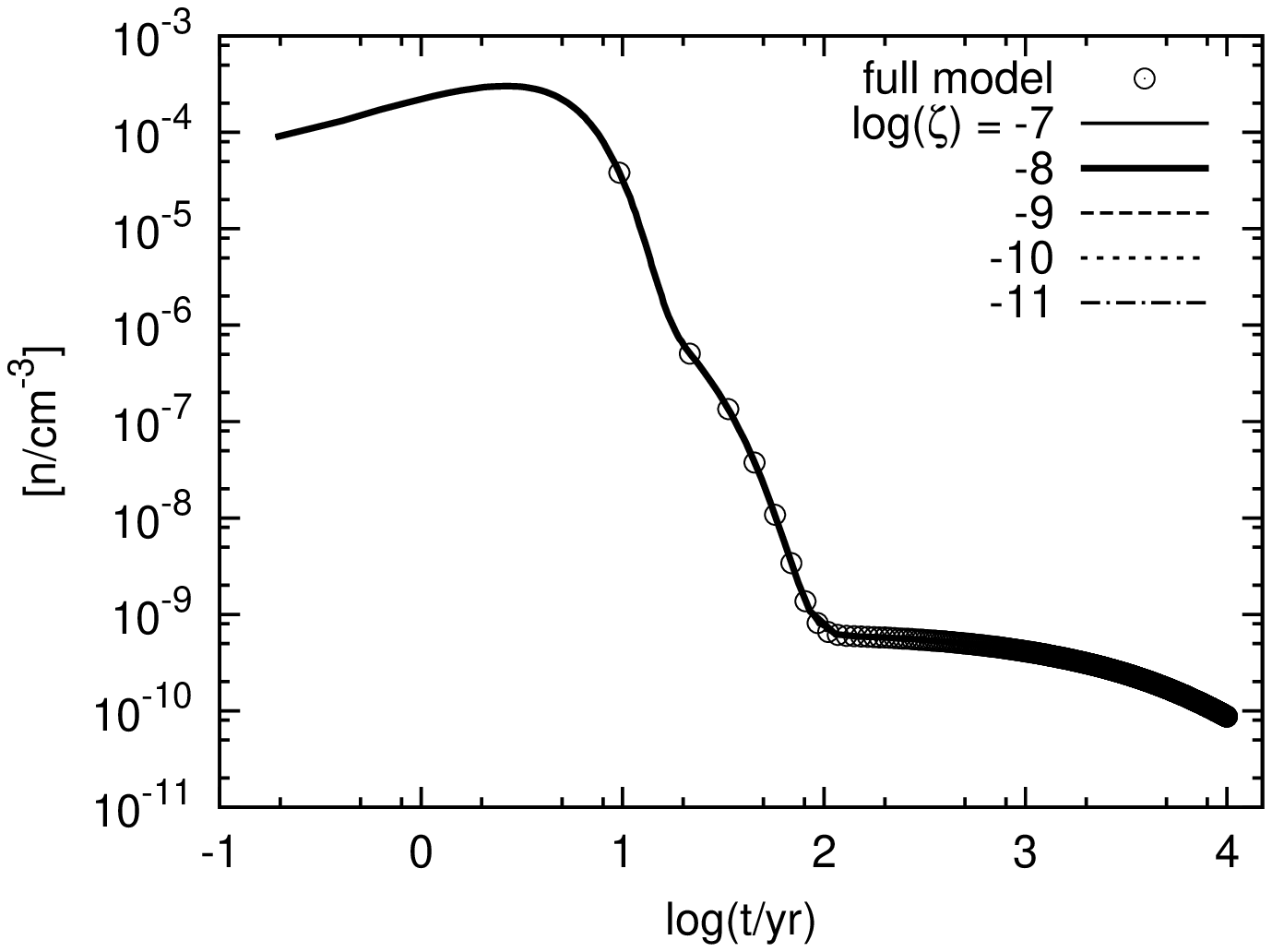}
	\includegraphics[width=.3\textwidth, angle=-90]{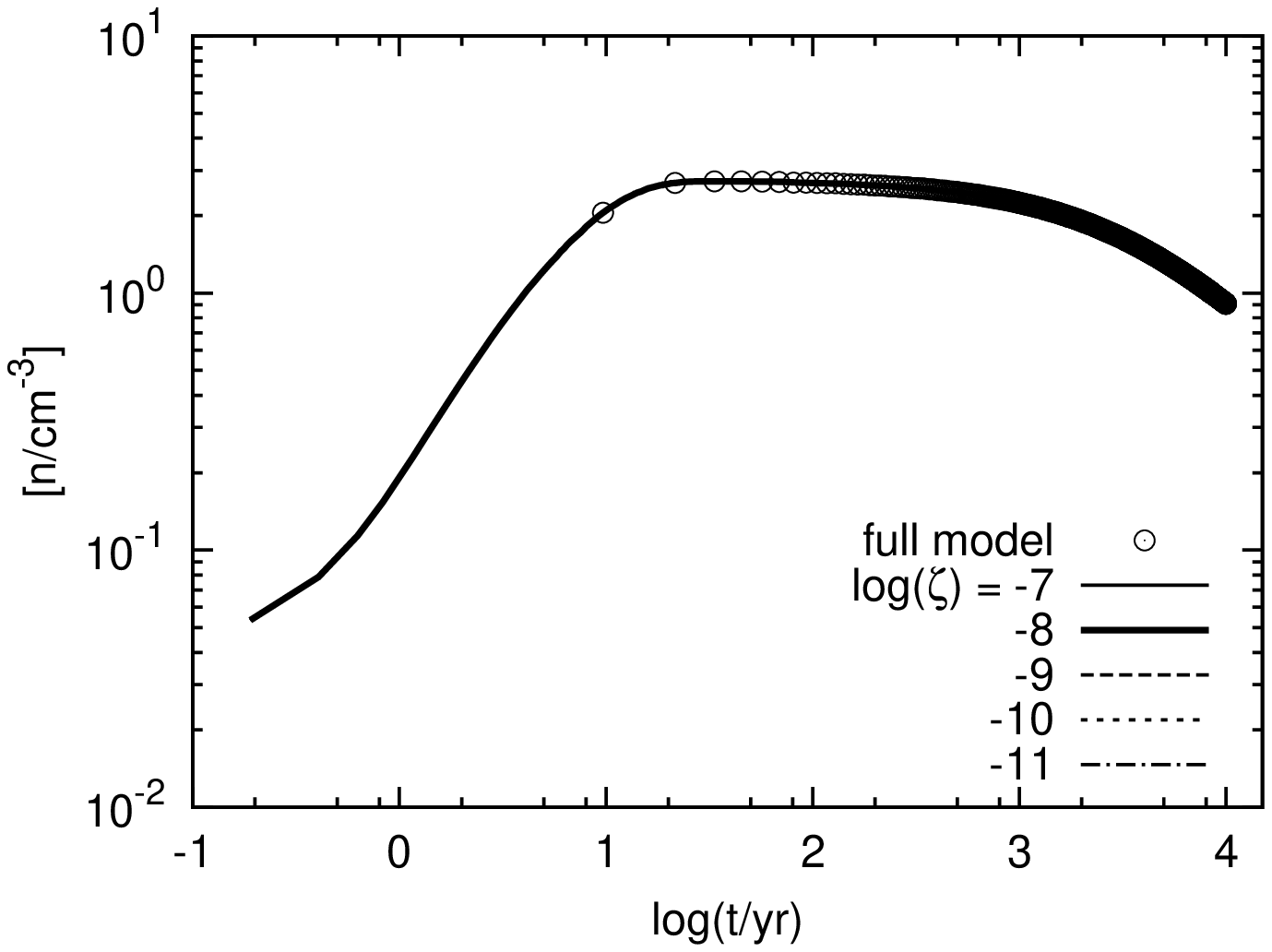}
	\caption{Large model calculations. Time evolutions of C (top) and Si (bottom) number densities. Open circles indicate the full model evolutions while lines are for different $\zeta$. Lines are clearly overlapping.}
	\label{wakC_Si}
\end{center}
\end{figure}

\begin{figure}
\begin{center}
	\includegraphics[width=.3\textwidth, angle=-90]{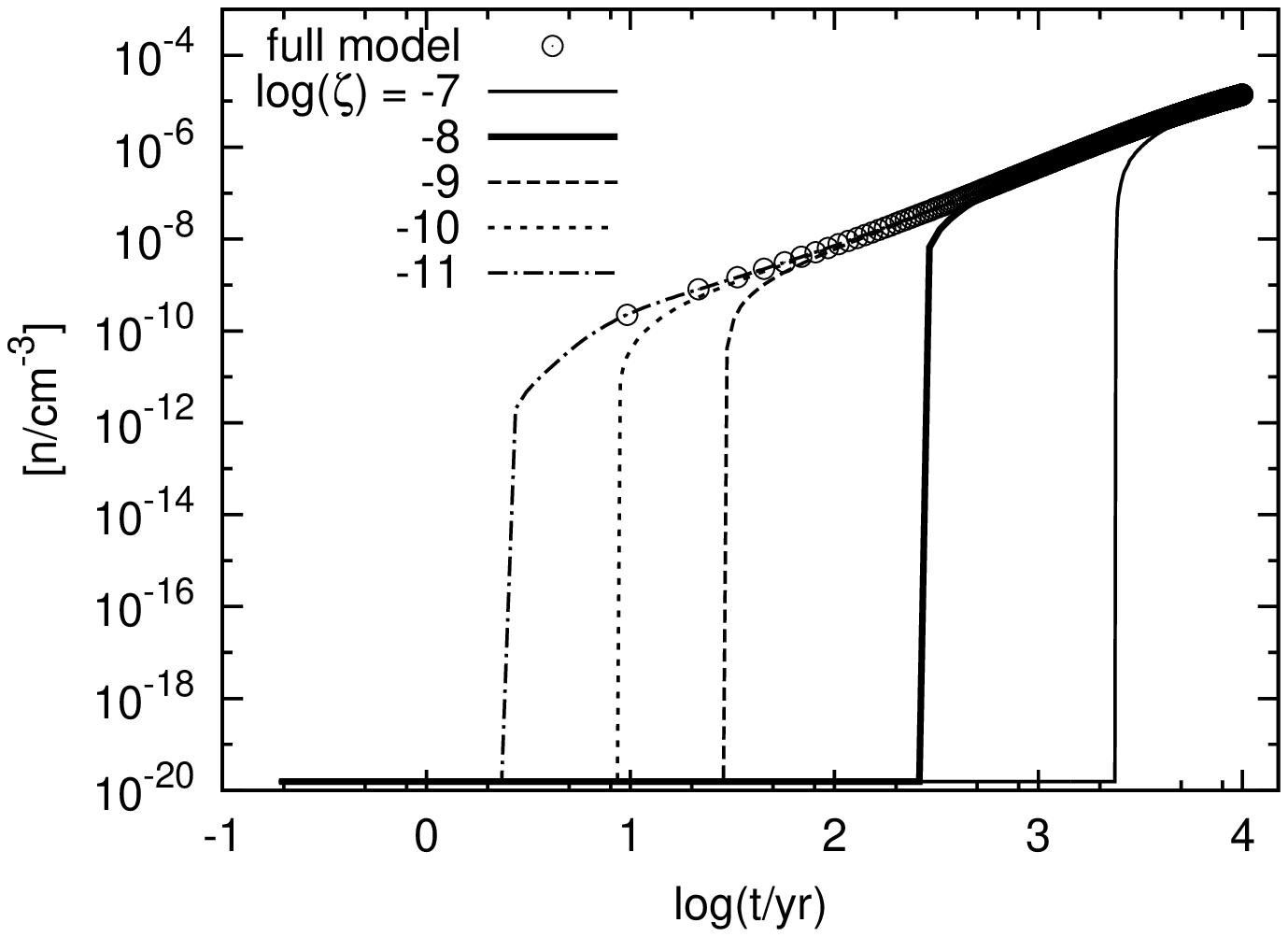}
	\includegraphics[width=.3\textwidth, angle=-90]{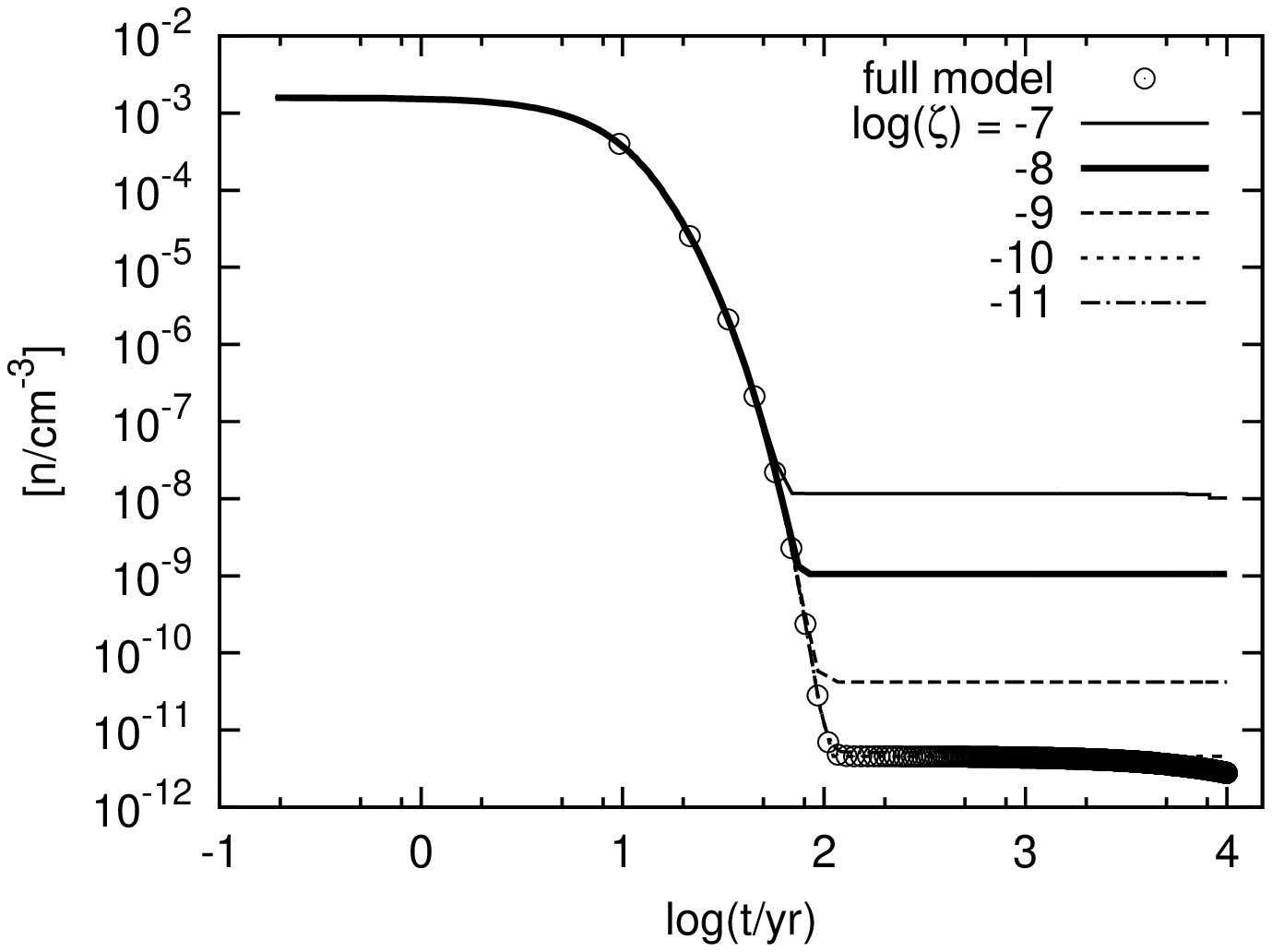}
	\caption{Large model calculations. Time evolutions of C$_3$S (top) and S$^+$ (bottom) number densities. Open circles indicate the full model evolutions while lines are for different $\zeta$.}
	\label{wakC3S_Sp}
\end{center}
\end{figure}

The results are similar to the ones found in the previous calculation. The time evolution of the various species can be divided into two classes: well-reproduced evolutions (as in Fig.\ref{wakC_Si}), and evolutions with a catch-up behaviour (Fig.\ref{wakC3S_Sp}). Note that in this latter case the model with $\zeta=10^{-11}$ almost matches the full model behaviour. Here, as in the small network case, the plots show a $\zeta$-depending shape for the same reasons previously described.

In this case we obtain a very large speed-up, since the RHS of Eq.\ref{ode_flux} is massively reduced (see Fig.\ref{glo_nrea}, bottom panel). The number of reactions (that depends on $\zeta$) grows in the first year and then decreases to lie on a long plateau. There is also more noise compared to the small model, because the length of the macro-steps is considerably smaller, as we see when the lines are around 1 yr where the system reaches its largest complexity and, thus, its smallest reduction.

To provide a table analogue to Tab.\ref{tabtfrac_glo1} an additional calculation for the \cite{Wakelam2008} network 
has been carried out. The table is available on request from the authors.

\section{Conclusions}
We have proposed a new network-reducing method obtained as an extension of that described by \cite{Tupper2002}. This reaction-based reduction approach allows one to increase the computational efficiency of solving the ODE system representing the chemical network. Our main modifications involve: (i) the introduction of an adaptive tolerance, (ii) a new definition of the macro-step length, as in Eq.(\ref{macrolength_abs}), and (iii) a sorting technique to obtain larger speed-ups.
We have applied our method to the chemistry of the ISM for a small network (similar to \citealt{Glover2010}), and for a larger one (similar to \citealt{Wakelam2008}). Our results show a very good agreement between the time-evolution of the full model and that of the reduced one, providing good speed-ups in the first case (about 50\%) and a very good one for the larger network (approximately 90\%). The accuracy and the speed-ups are coupled and their trade-off depends on the choice for the factor $\zeta$. 

The next step on further improvements could be that of introducing an intermediate check during the macro-step, in order to control the validity of the linear-growth assumption in the definition of $G_j$. This check allows one to avoid the errors that arise from a non-linear behaviour, and to verify the consistency of the macro-step length before the end of the step itself. This method could be considered an alternative to the macro-step rejection procedure in TUP02, with our scheme recovering the cpu-time which would be spent in the rejected macro-step. Our modifications should therefore provide a computational gain when strong non-linear behaviour occurs.  

It is worth noticing here that our approach, when compared to the other reducing techniques proposed over the years as already discussed in Section \ref{chemical_network}, is based on a robust mathematical background that ensures its applicability to a wide range of astrochemical problems and turns out to be very little problem-dependent.

This encouraging result suggests to apply this method to more complicated problems such as the three-dimensional hydrodynamical simulations of the system's evolution in the physical space within the ISM, where solving the ODEs involving the chemistry network along that simulation becomes even more cpu-demanding.

\section*{Acknowledgements}
We thank the Department of Astronomy of the Universit\`{a} degli Studi di Padova for providing us with the necessary computational facilities.
T.G. acknowledges the financial support from the PRIN project 2009-2013 and S.B. thanks the \mbox{CINECA} Consortium for the awarding of an \mbox{ISCRA} Research Fellowship. We are also grateful to the referee dr Valentine Wakelam for having noticed the error on the initial Si abundance.

\bibliographystyle{mn2e}      
\bibliography{mybib}

\begin{thebibliography}{}

\bibitem[\protect\citeauthoryear{{Galli} \& {Palla}}{{Galli} \&
  {Palla}}{1998}]{GalliPalla98}
{Galli} D.,  {Palla} F.,  1998, \aap, 335, 403

\bibitem[\protect\citeauthoryear{{Glover}, {Federrath}, {Mac Low} \&
  {Klessen}}{{Glover} et~al.}{2010}]{Glover2010}
{Glover} S.~C.~O.,  {Federrath} C.,  {Mac Low} M.-M.,    {Klessen} R.~S.,
  2010, \mnras, 404, 2

\bibitem[\protect\citeauthoryear{{Gnedin}, {Tassis} \& {Kravtsov}}{{Gnedin}
  et~al.}{2009}]{Gnedin2009}
{Gnedin} N.~Y.,  {Tassis} K.,    {Kravtsov} A.~V.,  2009, \apj, 697, 55

\bibitem[\protect\citeauthoryear{Hindmarsh}{Hindmarsh}{1983}]{Hindmarsh83}
Hindmarsh A.~C.,  1983, IMACS Transactions on Scientific Computation, 1, 55

\bibitem[\protect\citeauthoryear{{Knuth}}{{Knuth}}{1997}]{Knuth1997}
{Knuth} D.,  1997, The Art of Computer Programming, Volume 3: Sorting and
  Searching, 3rd edn.
Addison-Wesley, Boston, MA

\bibitem[\protect\citeauthoryear{{Maio}, {Dolag}, {Ciardi} \&
  {Tornatore}}{{Maio} et~al.}{2007}]{Maio07}
{Maio} U.,  {Dolag} K.,  {Ciardi} B.,    {Tornatore} L.,  2007, \mnras, 379,
  963

\bibitem[\protect\citeauthoryear{{Merlin} \& {Chiosi}}{{Merlin} \&
  {Chiosi}}{2007}]{MerlinChiosi07}
{Merlin} E.,  {Chiosi} C.,  2007, \aap, 473, 733

\bibitem[\protect\citeauthoryear{{Nelson} \& {Langer}}{{Nelson} \&
  {Langer}}{1999}]{Nelson1999}
{Nelson} R.~P.,  {Langer} W.~D.,  1999, \apj, 524, 923

\bibitem[\protect\citeauthoryear{Okino \& Mavrovouniotis}{Okino \&
  Mavrovouniotis}{1998}]{Okino98}
Okino M.~S.,  Mavrovouniotis M.~L.,  1998, Chemical Reviews, 98, 391

\bibitem[\protect\citeauthoryear{Petzold \& Zhu}{Petzold \&
  Zhu}{1999}]{Petzold1999}
Petzold L.,  Zhu W.,  1999, AIChE Journal, 45, 869

\bibitem[\protect\citeauthoryear{{Te{\c s}ileanu}, {Mignone} \&
  {Massaglia}}{{Te{\c s}ileanu} et~al.}{2008}]{Tesileanu2008}
{Te{\c s}ileanu} O.,  {Mignone} A.,    {Massaglia} S.,  2008, \aap, 488, 429

\bibitem[\protect\citeauthoryear{Tupper}{Tupper}{2002}]{Tupper2002}
Tupper P.~F.,  2002, Bit Numerical Mathematics, 42, 447

\bibitem[\protect\citeauthoryear{{Wakelam} \& {Herbst}}{{Wakelam} \&
  {Herbst}}{2008}]{Wakelam2008}
{Wakelam} V.,  {Herbst} E.,  2008, \apj, 680, 371

\bibitem[\protect\citeauthoryear{{Wiebe}, {Semenov} \& {Henning}}{{Wiebe}
  et~al.}{2003}]{Wiebe2003}
{Wiebe} D.,  {Semenov} D.,    {Henning} T.,  2003, \aap, 399, 197

\bibitem[\protect\citeauthoryear{{Yamasawa}, {Habe}, {Kozasa}, {Nozawa},
  {Hirashita}, {Umeda} \& {Nomoto}}{{Yamasawa} et~al.}{2011}]{Yamasawa2011}
{Yamasawa} D.,  {Habe} A.,  {Kozasa} T.,  {Nozawa} T.,  {Hirashita} H.,
  {Umeda} H.,    {Nomoto} K.,  2011, \apj, 735, 44

\end{thebibliography}

\appendix
\section{List of chemical reactions for the small network example}\label{appendixA}

The list of chemical reactions is reported in Tab. \ref{tabtfrac_glo1}.

\begin{table*}
	\centering
	\caption{Percentage of the total time that a given reaction participates in the reduced network with $\zeta=10^{-13}$. The values of the 
		reaction rates $k$ chosen for the small example are from \citet{Glover2010} except those involving carbon and 
		oxygen which are from OSU database (\osu). 
		This table refers to the small network example only.}
	\label{tabtfrac_glo1}
	\tiny
	\begin{tabular}{llclclclr|llclclclr}
		\hline
		\#&\multicolumn{7}{l}{Reaction}&$\%$&\#&\multicolumn{7}{l}{Reaction}&$\%$\\
		\hline
		1.&H&+&e$^-$&$\to$&H$^-$&+&$\gamma$&100.0&86.&H&+&CH$_3$$^+$&$\to$&CH$_2$$^+$&+&H$_2$&98.1\\
		2.&H$^-$&+&H&$\to$&H$_2$&+&e$^-$&100.0&87.&C$^+$&+&CH$_2$&$\to$&CH$_2$$^+$&+&C&97.8\\
		3.&H$^-$&+&H&$\to$&2H&+&e$^-$&100.0&88.&OH$^+$&+&CO&$\to$&HCO$^+$&+&O&97.4\\
		4.&H&+&CH&$\to$&C&+&H$_2$&100.0&89.&CH$_3$$^+$&+&e$^-$&$\to$&CH&+&H$_2$&97.4\\
		5.&H&+&CH$_2$&$\to$&CH&+&H$_2$&100.0&90.&CH$_2$&+&$\gamma$&$\to$&CH$_2$$^+$&+&e$^-$&97.4\\
		6.&H&+&OH&$\to$&O&+&H$_2$&100.0&91.&CH$_3$$^+$&+&e$^-$&$\to$&CH$_2$&+&H&96.1\\
		7.&H&+&O$_2$&$\to$&OH&+&O&100.0&92.&CH$_2$&+&$\gamma$&$\to$&CH&+&H&96.1\\
		8.&H$_2$&+&C&$\to$&CH&+&H&100.0&93.&H$^-$&+&H$^+$&$\to$&H$_2$$^+$&+&e$^-$&89.6\\
		9.&H$_2$&+&O&$\to$&OH&+&H&100.0&94.&H$^-$&+&H$^+$&$\to$&H$_2$$^+$&+&e$^-$&89.6\\
		10.&C&+&CH&$\to$&C$_2$&+&H&100.0&95.&O$^+$&+&OH&$\to$&O$_2$$^+$&+&H&89.6\\
		11.&C&+&OH&$\to$&CO&+&H&100.0&96.&H$^+$&+&H$_2$O&$\to$&H$_2$O$^+$&+&H&89.6\\
		12.&C&+&O$_2$&$\to$&CO&+&O&100.0&97.&O$^+$&+&OH&$\to$&OH$^+$&+&O&89.6\\
		13.&CH&+&O&$\to$&OH&+&C&100.0&98.&O$^+$&+&e$^-$&$\to$&O&+&$\gamma$&89.6\\
		14.&CH&+&O&$\to$&CO&+&H&100.0&99.&CH$_2$$^+$&+&$\gamma$&$\to$&CH&+&H$^+$&89.6\\
		15.&O&+&OH&$\to$&O$_2$&+&H&100.0&100.&CH$_2$$^+$&+&$\gamma$&$\to$&C$^+$&+&H$_2$&89.6\\
		16.&O&+&C$_2$&$\to$&CO&+&C&100.0&101.&CO&+&HOC$^+$&$\to$&HCO$^+$&+&CO&89.6\\
		17.&H&+&C$^-$&$\to$&CH&+&e$^-$&100.0&102.&H$_2$$^+$&+&e$^-$&$\to$&H&+&H&86.4\\
		18.&H&+&O$^-$&$\to$&OH&+&e$^-$&100.0&103.&OH$^+$&+&OH&$\to$&H$_2$O$^+$&+&O&86.4\\
		19.&H&+&C&$\to$&CH&+&$\gamma$&100.0&104.&H$^+$&+&C$^-$&$\to$&C&+&H&86.4\\
		20.&H&+&O&$\to$&OH&+&$\gamma$&100.0&105.&CH$_3$$^+$&+&$\gamma$&$\to$&CH$_2$$^+$&+&H&86.4\\
		21.&C&+&C&$\to$&C$_2$&+&$\gamma$&100.0&106.&CH$_3$$^+$&+&e$^-$&$\to$&CH$_3$&+&$\gamma$&85.2\\
		22.&C&+&O&$\to$&CO&+&$\gamma$&100.0&107.&H&+&CH$_3$&$\to$&CH$_2$&+&H$_2$&83.6\\
		23.&O&+&O&$\to$&O$_2$&+&$\gamma$&100.0&108.&CH$_2$$^+$&+&O$_2$&$\to$&HCO$^+$&+&OH&54.0\\
		24.&C&+&$\gamma$&$\to$&C$^+$&+&e$^-$&100.0&109.&H$_2$O$^+$&+&CO&$\to$&HCO$^+$&+&OH&54.0\\
		25.&CH&+&$\gamma$&$\to$&C&+&H&100.0&110.&OH$^+$&+&O$_2$&$\to$&O$_2$$^+$&+&OH&54.0\\
		26.&CH&+&O&$\to$&HCO$^+$&+&e$^-$&100.0&111.&H$_2$&+&C$^-$&$\to$&CH$_2$&+&e$^-$&54.0\\
		27.&C&+&OH&$\to$&O&+&CH&100.0&112.&OH$^+$&+&$\gamma$&$\to$&O$^+$&+&H&54.0\\
		28.&H$_2$&+&C&$\to$&CH$_2$&+&$\gamma$&100.0&113.&CH&+&HCO$^+$&$\to$&CO&+&CH$_2$$^+$&46.0\\
		29.&OH&+&$\gamma$&$\to$&O&+&H&100.0&114.&H$_2$&+&H$^+$&$\to$&H$_2$$^+$&+&H&6.5\\
		30.&H$_2$&+&CH&$\to$&CH$_2$&+&H&100.0&115.&H$_2$&+&CH$_2$&$\to$&CH$_3$&+&H&6.5\\
		31.&C&+&e$^-$&$\to$&C$^-$&+&$\gamma$&100.0&116.&H$_2$&+&O$_2$&$\to$&OH&+&OH&6.5\\
		32.&O&+&e$^-$&$\to$&O$^-$&+&$\gamma$&100.0&117.&H$^+$&+&CH$_2$&$\to$&CH$_2$$^+$&+&H&6.5\\
		33.&H$_2$&+&OH&$\to$&H$_2$O&+&H&100.0&118.&H$^+$&+&CH$_3$&$\to$&CH$_3$$^+$&+&H&6.5\\
		34.&OH&+&OH&$\to$&H$_2$O&+&O&100.0&119.&O$^+$&+&H$_2$O&$\to$&H$_2$O$^+$&+&O&6.5\\
		35.&OH&+&$\gamma$&$\to$&OH$^+$&+&e$^-$&100.0&120.&O$^+$&+&O$_2$&$\to$&O$_2$$^+$&+&O&6.5\\
		36.&C$_2$&+&$\gamma$&$\to$&C&+&C&100.0&121.&OH$^+$&+&H$_2$O&$\to$&H$_2$O$^+$&+&OH&6.5\\
		37.&CO&+&$\gamma$&$\to$&O&+&C&100.0&122.&H$_2$O$^+$&+&O$_2$&$\to$&O$_2$$^+$&+&H$_2$O&6.5\\
		38.&O$_2$&+&$\gamma$&$\to$&O&+&O&100.0&123.&C$^-$&+&O$^+$&$\to$&O&+&C&6.5\\
		39.&O&+&OH$^+$&$\to$&O$_2$$^+$&+&H&100.0&124.&CH$_3$&+&$\gamma$&$\to$&CH$_3$$^+$&+&e$^-$&6.5\\
		40.&OH&+&HCO$^+$&$\to$&CO&+&H$_2$O$^+$&100.0&125.&CH$_3$&+&$\gamma$&$\to$&CH$_2$&+&H&6.5\\
		41.&HCO$^+$&+&e$^-$&$\to$&CO&+&H&100.0&126.&CH$_3$&+&$\gamma$&$\to$&CH&+&H$_2$&6.5\\
		42.&O$_2$&+&$\gamma$&$\to$&O$_2$$^+$&+&e$^-$&100.0&127.&H$_2$O$^+$&+&$\gamma$&$\to$&OH$^+$&+&H&6.5\\
		43.&C$^-$&+&O&$\to$&CO&+&e$^-$&100.0&128.&H$_2$&+&e$^-$&$\to$&2H&+&H&0.0\\
		44.&C&+&O$^-$&$\to$&CO&+&e$^-$&100.0&129.&H$_2$&+&H&$\to$&2H&+&H&0.0\\
		45.&C$^+$&+&O$_2$&$\to$&CO&+&O$^+$&100.0&130.&H$_2$&+&H$_2$&$\to$&H$_2$&+&2H&0.0\\
		46.&C$^-$&+&C&$\to$&C$_2$&+&e$^-$&100.0&131.&H&+&e$^-$&$\to$&H$^+$&+&2e$^-$&0.0\\
		47.&O$^-$&+&O&$\to$&O$_2$&+&e$^-$&100.0&132.&H&+&C$_2$&$\to$&CH&+&C&0.0\\
		48.&H&+&O$^+$&$\to$&O&+&H$^+$&100.0&133.&H&+&CO&$\to$&OH&+&C&0.0\\
		49.&H&+&H$_2$O&$\to$&OH&+&H$_2$&100.0&134.&C&+&CH$_2$&$\to$&CH&+&CH&0.0\\
		50.&C$^+$&+&e$^-$&$\to$&C&+&$\gamma$&100.0&135.&C&+&CO&$\to$&C$_2$&+&O&0.0\\
		51.&H$^+$&+&O&$\to$&O$^+$&+&H&100.0&136.&CH&+&O&$\to$&CO&+&H&0.0\\
		52.&CH$_2$$^+$&+&O&$\to$&HCO$^+$&+&H&100.0&137.&CH&+&O$_2$&$\to$&CO&+&OH&0.0\\
		53.&C&+&O$_2$$^+$&$\to$&O$_2$&+&C$^+$&100.0&138.&CH$_2$&+&CH$_2$&$\to$&CH$_3$&+&CH&0.0\\
		54.&O&+&H$_2$O$^+$&$\to$&O$_2$$^+$&+&H$_2$&99.9&139.&CH$_2$&+&O&$\to$&OH&+&CH&0.0\\
		55.&H$_2$&+&C$^+$&$\to$&CH$_2$$^+$&+&$\gamma$&99.9&140.&CH$_2$&+&O&$\to$&CO&+&H$_2$&0.0\\
		56.&O$^-$&+&$\gamma$&$\to$&O&+&e$^-$&99.9&141.&CH$_2$&+&OH&$\to$&O&+&CH$_3$&0.0\\
		57.&C$^-$&+&$\gamma$&$\to$&C&+&e$^-$&99.9&142.&CH$_2$&+&OH&$\to$&H$_2$O&+&CH&0.0\\
		58.&H&+&H$^+$&$\to$&H$_2$$^+$&+&$\gamma$&99.7&143.&CH$_2$&+&O$_2$&$\to$&CO&+&H$_2$O&0.0\\
		59.&O&+&H$_2$O&$\to$&OH&+&OH&99.7&144.&CH$_3$&+&OH&$\to$&H$_2$O&+&CH$_2$&0.0\\
		60.&H$_2$&+&OH$^+$&$\to$&H$_2$O$^+$&+&H&99.7&145.&O&+&C$_2$&$\to$&CO&+&C&0.0\\
		61.&C$^+$&+&H$_2$O&$\to$&HOC$^+$&+&H&99.7&146.&C$_2$&+&O$_2$&$\to$&CO&+&CO&0.0\\
		62.&C$^+$&+&C$^-$&$\to$&C&+&C&99.7&147.&O&+&OH&$\to$&O$_2$&+&H&0.0\\
		63.&OH$^+$&+&e$^-$&$\to$&O&+&H&99.7&148.&H$_3$$^+$&+&CH&$\to$&CH$_2$$^+$&+&H$_2$&0.0\\
		64.&O$_2$$^+$&+&e$^-$&$\to$&O&+&O&99.7&149.&H$_3$$^+$&+&CH$_2$&$\to$&CH$_3$$^+$&+&H$_2$&0.0\\
		65.&H$_2$O&+&$\gamma$&$\to$&OH&+&H&99.7&150.&H$_3$$^+$&+&O&$\to$&OH$^+$&+&H$_2$&0.0\\
		66.&H&+&H$_2$$^+$&$\to$&H$_2$&+&H$^+$&99.5&151.&H$_3$$^+$&+&O&$\to$&H$_2$O$^+$&+&H&0.0\\
		67.&C$^+$&+&H$_2$O&$\to$&HCO$^+$&+&H&99.5&152.&H$_3$$^+$&+&OH&$\to$&H$_2$O$^+$&+&H$_2$&0.0\\
		68.&H$^+$&+&OH&$\to$&OH$^+$&+&H&99.5&153.&H$_3$$^+$&+&CO&$\to$&HCO$^+$&+&H$_2$&0.0\\
		69.&H$_2$O$^+$&+&e$^-$&$\to$&O&+&H$_2$&99.5&154.&H$_3$$^+$&+&CO&$\to$&HOC$^+$&+&H$_2$&0.0\\
		70.&H$_2$O$^+$&+&e$^-$&$\to$&OH&+&H&99.5&155.&C$^-$&+&O$_2$&$\to$&CO&+&O$^-$&0.0\\
		71.&H$_2$O&+&$\gamma$&$\to$&H$_2$O$^+$&+&e$^-$&99.5&156.&CH&+&OH$^+$&$\to$&O&+&CH$_2$$^+$&0.0\\
		72.&CH$_2$$^+$&+&e$^-$&$\to$&CH&+&H&99.4&157.&CH&+&H$_2$O$^+$&$\to$&OH&+&CH$_2$$^+$&0.0\\
		73.&H$_2$&+&O$^-$&$\to$&H$_2$O&+&e$^-$&99.4&158.&CH&+&O$_2$$^+$&$\to$&HCO$^+$&+&O&0.0\\
		74.&O$_2$$^+$&+&$\gamma$&$\to$&O$^+$&+&O&99.4&159.&CH$_2$&+&OH$^+$&$\to$&O&+&CH$_3$$^+$&0.0\\
		75.&H$_2$&+&O$^+$&$\to$&OH$^+$&+&H&99.0&160.&CH$_2$&+&H$_2$O$^+$&$\to$&OH&+&CH$_3$$^+$&0.0\\
		76.&H$^+$&+&O$_2$&$\to$&O$_2$$^+$&+&H&99.0&161.&CH$_2$&+&HCO$^+$&$\to$&CO&+&CH$_3$$^+$&0.0\\
		77.&CH$_2$$^+$&+&e$^-$&$\to$&C&+&H$_2$&99.0&162.&H$^+$&+&O&$\to$&O$^+$&+&H&0.0\\
		78.&HOC$^+$&+&e$^-$&$\to$&CO&+&H&99.0&163.&CH$_2$&+&O$^+$&$\to$&O&+&CH$_2$$^+$&0.0\\
		79.&H$^+$&+&e$^-$&$\to$&H&+&$\gamma$&99.0&164.&CH$_2$&+&OH$^+$&$\to$&OH&+&CH$_2$$^+$&0.0\\
		80.&H$^-$&+&e$^-$&$\to$&H&+&2e$^-$&98.7&165.&CH$_2$&+&H$_2$O$^+$&$\to$&H$_2$O&+&CH$_2$$^+$&0.0\\
		81.&H$_2$&+&CH$_2$$^+$&$\to$&CH$_3$$^+$&+&H&98.7&166.&CH$_2$&+&O$_2$$^+$&$\to$&O$_2$&+&CH$_2$$^+$&0.0\\
		82.&H$^-$&+&H$^+$&$\to$&H&+&H&98.2&167.&H$_3$$^+$&+&e$^-$&$\to$&H$_2$&+&H&0.0\\
		83.&H$^+$&+&e$^-$&$\to$&H&+&$\gamma$&98.2&168.&H&+&OH&$\to$&H$_2$O&+&$\gamma$&0.0\\
		84.&CH$_3$$^+$&+&O&$\to$&HCO$^+$&+&H$_2$&98.2&169.&H$_2$&+&CH&$\to$&CH$_3$&+&$\gamma$&0.0\\
		85.&H$_2$&+&HOC$^+$&$\to$&HCO$^+$&+&H$_2$&98.2&170.&H$_3$$^+$&+&$\gamma$&$\to$&H$_2$&+&H$^+$&0.0\\
		\hline
	\end{tabular}
\end{table*}

\bsp

\label{lastpage}

\end{document}